\documentclass[12pt, onecolumn, draftcls]{IEEEtran}

\usepackage{amsmath}
\usepackage{amsfonts}
\usepackage{amssymb}
\usepackage{graphicx}
\usepackage{epsfig}
\usepackage{mathrsfs}
\usepackage{cite}
\usepackage{psfig}
\usepackage{epsf}
\usepackage{flushend}
\usepackage[active]{srcltx}

\newtheorem{theorem}{Theorem}

\newtheorem{lemma}[theorem]{Lemma}
\newtheorem{corol}[theorem]{Corollary}

\begin{document}

\title{Throughput Scaling Laws for Wireless Networks with Fading Channels}
\author{Masoud Ebrahimi, Mohammad A. Maddah-Ali, and Amir K. Khandani
\thanks{This work is financially supported by Nortel Networks and by matching funds from
the federal government of Canada (NSERC) and province of Ontario
(OCE).}
\thanks{The authors are affiliated with the Coding and Signal Transmission Laboratory,
Electrical and Computer Engineering Department, University of
Waterloo, Waterloo, ON, N2L 3G1, Canada, Tel: 519-885-1211 ext.
35324, \mbox{Fax: 519-888-4338}, Emails:~ \{masoud, mohammad,
khandani\}@cst.uwaterloo.ca.}
\thanks{This material was presented in part at the 40th Conference on Information Sciences and
Systems (CISS), Princeton University, Princeton, NJ, March
2006.}}\maketitle \markboth{Submitted to IEEE Transactions on
Information Theory}{}

\begin{abstract}
A network of $n$ communication links, operating over a shared
wireless channel, is considered. Fading is assumed to be the
dominant factor affecting the strength of the channels between
transmitter and receiver terminals. It is assumed that each link can
be active and transmit with a constant power $P$ or remain silent.
The objective is to maximize the throughput over the selection of
active links. By deriving an upper bound and a lower bound, it is
shown that in the case of Rayleigh fading (i) the maximum throughput
scales like $\log n$ (ii) the maximum throughput is achievable in a
distributed fashion. The upper bound is obtained using probabilistic
methods, where the key point is to upper bound the throughput of any
random set of active links by a chi-squared random variable. To
obtain the lower bound, a decentralized link activation strategy is
proposed and analyzed.
\end{abstract}


\section{Introduction}
In a wireless network, a number of source nodes transmit data to
their designated destination nodes through a shared wireless
channel. Analysis and design of such configurations, even in their
simplest forms, have been among the most difficult problems facing
the network information theory community for many years.

Followed by the pioneering work of Gupta and Kumar \cite{Gupta00},
considerable attention has been paid to investigate how the
throughput of wireless networks scales with $n$, the number of
nodes, when $n$ is large. This has been done assuming different
network topologies, traffic patterns, protocol schemes, and channel
models \cite{Gupta00, Gastpar05, Xie04, Leveque05, Franceschetti07j,
Grossglauser02, Kulkarni04, Toumpis04, Xue05, Gowaikar05}. Most of
these works consider a channel model in which the signal power
decays according to a distance-based attenuation law
\cite{Gupta00,Franceschetti07j,Xie04,Gastpar05,Leveque05,Grossglauser02,Kulkarni04}.
However, in a wireless environment the presence of obstacles and
scatterers adds some randomness to the received signal. This random
behavior of the channel, known as fading, can drastically change the
scaling laws of a network in both multihop
\cite{Gowaikar05,Toumpis04,Xue05} and single-hop scenarios
\cite[Chapter 8]{Etkinthesis}, \cite{Weber06, Gesbert07rawnet}.

In this paper, we follow the model of \cite{Gowaikar05,
Etkinthesis}, where fading is assumed to be the dominant factor
affecting the strength of the channels between nodes. Despite the
randomness of the channel, we are only interested in events that
occur asymptotically almost surely, i.e., with probability tending
to one as $n \to \infty$. Such a deterministic approach to random
wireless networks has been also adopted in
\cite{Kulkarni04,Franceschetti07j}, where the nodes' locations are
random.

We consider a single-hop scenario, i.e., a network structure in
which the transmitters send data to their corresponding receivers
directly and without utilizing other nodes as routers. It is assumed
that each link can be active and transmit with a constant power $P$
or remain silent. The objective is to maximize the throughput over
all sets of active links. We propose a threshold-based link
activation strategy in which each link is active if and only if its
channel gain is above some predetermined threshold. The decision on
being active can be made at the receivers, where their own channel
gains are estimated and a single-bit command data is fed back to the
transmitters. Hence, there is no need for the exchange of
information between links. Consequently, this method can be
implemented in a decentralized fashion. We analyze this method for a
general fading model and show how to obtain the value of the
activation threshold to maximize the throughput. As an example, we
derive a closed form expression for the achievable throughput in the
Rayleigh fading environment.

Using probabilistic methods, we derive an upper bound on the
achievable throughput when the channel is Rayleigh fading.
Interestingly, this upper bound scales the same as the lower bound
achieved by the proposed strategy. This proves the asymptotic
optimality of the proposed technique among all link activation
strategies.

In addition to the channel modeling, \cite{Gowaikar05} is a relevant
work in the sense that transmissions occur with the same power and
the objective is to maximize the throughput. However, they allow
multihop communication in their scheme. Their proposed scheme
requires a central unit which is aware of all channel conditions and
decides on active source-destination pairs and the paths between
them. Despite this complexity, the achievable throughput of their
method in the popular model of Rayleigh fading is by a factor of 4
less than the value obtained in this work for a more restricted
configuration, i.e., single-hop networks with decentralized
management\footnote{Note that the method proposed in
\cite{Gowaikar05} is general and can be applied for any fading
distribution. Here, we have made the comparison just for the
Rayleigh fading model. For the comparison in other fading
environments, see \cite{MyTechnical2}.}.

The rest of the paper is organized as follows: In Section
\ref{system}, the network model and problem formulation are
presented. By proposing a decentralized link activation strategy, a
lower bound on the network throughput is derived in Section
\ref{achievable}. In Section~\ref{UTMoptimal}, we prove the
optimality of the proposed decentralized method in a Rayleigh fading
environment. Finally, we conclude the paper in
Section~\ref{conclusion}.

\emph{Notation:} $\mathcal{N}_n$ represents the set of natural
numbers less than or equal to $n$; $\log ( \cdot )$ is the natural
logarithm function; $\textrm{P} (A)$ denotes the probability of
event $A$; $\textrm{E}(x)$ represents the expected value of the
random variable $x$; $\approx$ means approximate equality; for any
functions $f(n)$ and $h(n)$, \mbox{$h(n)=O(f(n))$} is equivalent to
$ \lim_{n \to \infty} \left | {h(n)}/ {f(n)} \right |< \infty$,
\mbox{$h(n)=o(f(n))$} is equivalent to $ \lim_{n \to \infty} \left |
{h(n)}/{f(n)} \right | =0$, \mbox{$h(n)=\omega(f(n))$} is equivalent
to $ \lim_{n \to \infty} \left | {h(n)}/{f(n)} \right | = \infty$,
and \mbox{$h(n)\sim f(n)$} is equivalent to $ \lim_{n \to \infty}
{h(n)}/{f(n)}  = 1$; an event $A_n$ holds \emph{asymptotically
almost surely} (a.a.s) if $\textrm{P} (A_n) \to 1$ as $ n \to
\infty$.

\section{Network Model and Problem Formulation}\label{system}

We consider a wireless communication network with $n$ pairs of
transmitters and receivers. These $n$ communication links are
indexed by the elements of $\mathcal{N}_n$. Each transmitter aims to
send data to its corresponding receiver in a single-hop fashion. The
transmit power of link $i$ is denoted by $p_i$. It is assumed that
the links follow an on-off paradigm, i.e., $p_i \in \{ 0,\,P \} $,
where $P$ is a constant. Hence, any power allocation scheme
translates to a \emph{link activation strategy (LAS)}. Any LAS
yields a set of active links $\mathcal{A}$, which describes the
transmission powers as
\begin{equation}
p_i=\left \{
\begin{tabular}{lcl}
$P$ & if & $i \in \mathcal{A} $\\
$0$ & if & $i \notin \mathcal{A}$
\end{tabular}.
\right .
\end{equation}

The channel between transmitter $j$ and receiver $i$ is
characterized by the coefficient $g_{ji}$. This means the received
power from transmitter $j$ at the receiver $i$ equals $g_{ji}p_j$.
We assume that the channel coefficients are \emph{independent
identically distributed} (i.i.d.) random variables drawn from a pdf
$f(x)$ with mean $\mu$ and variance $\sigma^2$. The channel between
transmitter $i$ and receiver $i$ is simply referred to as the
\emph{direct channel} of link $i$.

We consider an additive white Gaussian noise (AWGN) with limited
variance $\eta$ at the receivers. The transmit \emph{signal-to-noise
ratio} (\emph{SNR}) of the network is defined as
\begin{equation}
\rho=\dfrac{P}{\eta}.
\end{equation}
The receivers are conventional linear receivers, i.e., without
multiuser detection. Since the transmissions occur simultaneously
within the same environment, the signal from each transmitter acts
as interference for other links. Assuming Gaussian signal
transmission from all links, the distribution of the interference
will be Gaussian as well. Thus, according to the Shannon capacity
formula, the maximum supportable rate of link $i \in \mathcal{A}$ is
obtained as
\begin{equation}\label{ratedef}
r_i (\mathcal{A})=\log \left( 1+ \gamma_i(\mathcal{A}) \right) \;
\textrm{nats/channel use},
\end{equation}
where
\begin{equation}\label{SINRdef}
\gamma_i(\mathcal{A})=\dfrac{g_{ii}}{1/\rho+\sum_{\substack{j \in \mathcal{A}\\
j \neq i}} g_{ji}}
\end{equation}
is the \emph{signal-to-interference-plus-noise ratio} (\emph{SINR})
of link $i$.

As a measure of performance, in this paper we consider the
throughput of the network, which is defined as
\begin{equation}\label{throughputdef}
T(\mathcal{A})=\sum_{i \in \mathcal{A}} r_i(\mathcal{A}).
\end{equation}
Also, the average rate per active link, or simply rate-per-link, is
defined as
\begin{equation}
\bar{r}(\mathcal{A})=\dfrac{T(\mathcal{A})}{|\mathcal{A}|}.
\end{equation}
In this paper, wherever there is no ambiguity, we drop the
functionality of $\mathcal{A}$ from the network parameters and
simply refer to them as $r_i$, $\gamma_i$, $T$, or $\bar{r}$.

The problem of throughput maximization is described as
\begin{eqnarray}\label{unconstrainedproblem}
\max_{\mathcal{A} \subseteq \mathcal{N}_n} & & T(\mathcal{A}).
\end{eqnarray}

We denote the maximum value of this problem by $T^*$. Due to the
nonconvex and integral nature of the throughput maximization
problem, its solution is computationally intensive. However, in this
paper we propose and analyze a decentralized LAS which leads to
efficient solutions for the above problem. Indeed, we show that the
proposed strategy a.a.s. achieves the optimum solution of the
throughput maximization problem in Rayleigh fading environment.

\section{Achievability Result}\label{achievable}

In this section, to derive a lower bound on the network throughput,
we propose a simple heuristic LAS, which we call a
\emph{threshold-based LAS (TBLAS)}. Due to the randomness of the
channel, the achievable throughput of the proposed strategy is a
random variable; however, our analysis yields a deterministic lower
bound which is a.a.s. achievable.

\emph{TBLAS:} For a threshold $\Delta$, choose the set of active
links according to the following rule
\begin{equation}\label{strategy1}
i \in \mathcal{A} \quad \textrm{iff} \quad g_{ii} > \Delta.
\end{equation}

If each transmitter is aware of the threshold $\Delta$ and its
direct channel coefficient, it can individually determine its
transmit power. Hence, TBLAS can be implemented in a
\emph{decentralized} fashion. To obtain the optimum value of
$\Delta$, we should first know the achievable throughput of TBLAS in
terms of $\Delta$.

Let $k_{\Delta}=|\mathcal{A}|$ denote the number of active links
chosen by TBLAS with a threshold $\Delta$. Without loss of
generality, we assume that
$\mathcal{A}=\{ 1,\,2,\,\cdots,\,k_{\Delta} \}$. By defining $I_i= \sum_{\substack{j=1\\
j \neq i}}^{k_{\Delta}} g_{ji}$ and using (\ref{ratedef}),
(\ref{SINRdef}), and (\ref{throughputdef}), the throughput can be
lower bounded as
\begin{eqnarray}
T  & > & \sum_{i=1}^{k_{\Delta}} \log \left(
1+\dfrac{\Delta}{1/\rho+I_i}
\right ) \label{lbound2} \\
& \geq & k_{\Delta}  \log \left( 1+\dfrac{\Delta}{
1/\rho+\frac{1}{k_{\Delta}} \sum_{i=1}^{k_{\Delta}} I_i} \right ),
\label{lbound3}
\end{eqnarray}
where the first equality is based on the fact that $g_{ii}>\Delta$
for the active links and the second one is the result of applying
the Jensen's inequality.

To simplify the RHS of (\ref{lbound3}), we apply the Chebyshev
inequality to obtain the upper bound
\begin{equation}\label{iupperbound}
\frac{1}{k_{\Delta}} \sum_{i=1}^{k_{\Delta}} I_i <(k_{\Delta}-1)\mu
+ \psi,
\end{equation}
which holds a.a.s. for any $\psi=\omega(1)$. Consequently, the lower
bound (\ref{lbound3}) becomes
\begin{equation}\label{lbound7}
T   > k_{\Delta}  \log \left( 1+\dfrac{\Delta}{\mu k_{\Delta} +
\psi} \right ),\quad a.a.s.
\end{equation}
Note that the constant $1/\rho - \mu$ is absorbed in the function
$\psi$. Let $q_{\Delta}$ denote the probability of a link being
active. We have $q_{\Delta}=1-F(\Delta)$, where $F(x)$ is the
cumulative distribution function (cdf) of the channel gains. Due to
the TBLAS, the number of active links, $k_{\Delta}$, is a binomial
random variable with parameters $n$ and $q_{\Delta}$. Using the
Chebyshev inequality, it can be shown that $k_{\Delta}$ a.a.s.
satisfies the lower bound
\begin{equation}\label{klowerbound}
k_{\Delta}> nq_{\Delta} - \xi \sqrt {nq_{\Delta}},
\end{equation}
for any $\xi=\omega(1)$. Noting that for $\psi=o(k_{\Delta})$, the
lower bound (\ref{lbound7}) becomes an increasing function of
$k_{\Delta}$, and by using (\ref{klowerbound}), we obtain the main
result of this section, which is an achievability result on the
throughput.

\begin{theorem}\label{achievablesumrate}
Consider a wireless network with $n$ links and i.i.d. random channel
coefficients with pdf $f(x)$, cdf $F(x)$, and mean $\mu$. Choose any
$\Delta>0$ and define \mbox{$q_{\Delta}=1-F(\Delta)$}. Then, a
throughput of
\begin{equation}\label{lbound6}
T_a (\Delta) = (nq_{\Delta} - \xi \sqrt {nq_{\Delta}}) \log \left (
1 + \dfrac{\Delta}{\mu(nq_{\Delta} - \xi \sqrt {nq_{\Delta}}) +\psi
} \right )
\end{equation}
\sloppy is a.a.s. achievable for any $\xi=\omega(1)$ that satisfies
$\xi=o(\sqrt{nq_{\Delta}})$ and any $\psi=\omega(1)$ that satisfies
$\psi=o(nq_{\Delta})$.
\end{theorem}
\fussy

Note that the achievable throughput $T_a (\Delta)$ is a
deterministic value. It easily follows that under the conditions
described in Theorem~\ref{achievablesumrate}, the number of active
links and the achievable average rate-per-link in TBLAS scale as
\cite{MyTechnical2}
\begin{eqnarray}
k_{\Delta} & \sim & nq_{\Delta} \quad a.a.s. \label{nomberofusers} \\
\bar{r}_{\Delta} & \sim &   \log \left ( 1 +
\dfrac{\Delta}{\mu(nq_{\Delta} - \xi \sqrt {nq_{\Delta}}) +\psi }
\right ) \quad a.a.s.
\end{eqnarray}

As specified in Theorem \ref{achievablesumrate}, the achievable
throughput of TBLAS is a function of the parameter $\Delta$. Thus,
$\Delta$ can be chosen such that the achievable throughput is
maximized. Let us define
\begin{equation}\label{optimumdelta}
\Delta^* = \arg \max_{\Delta} T_a (\Delta),
\end{equation}
and
\begin{equation}\label{optimumsumrate}
T_a ^*= \max_{\Delta} T_a (\Delta).
\end{equation}
In the following example,we clarify how to obtain these values for
the popular Rayleigh fading model.

\emph{Example:} In a Rayleigh fading channel, $f(x)=e^{-x}$,
$\mu=1$, and $q_{\Delta}=e^{-\Delta}$. By substituting $q_{\Delta}$
in (\ref{lbound6}), we obtain
\begin{equation}\label{originalobjective}
\!T_a (\Delta)\!= \!\! \left ( n e^{-\Delta} - \xi \sqrt{ n
e^{-\Delta}} \right ) \log \! \left ( \! 1+ \dfrac{\Delta}{n
e^{-\Delta} - \xi \sqrt{ n e^{-\Delta}}+\psi} \right ).
\end{equation}
The result of maximizing this function over $\Delta$ is given in the
following corollary.

\begin{corol}\label{optimumthresholdrayleigh}
Assuming Rayleigh fading, we have
\begin{eqnarray}
\Delta^* & = & \log n-2\log \log n + \log 2 +  O \left(
\dfrac{\log \log n}{\log n} \right), \label{rayleightopt} \\
{T_a} ^* & = & \log n -2  \log \log n + \log (2/e) + O \left(
\dfrac{\log \log n}{\log n} \right) , \; \; a.a.s. \label{maxthroughputrayleigh1}\\
k_{\Delta^*} & = &  \frac{1}{2} \log ^2 n (1+o(1)), \quad a.a.s., \\
\bar{r}_{\Delta^*}  & = &    \dfrac{2}{\log n} (1+o(1)), \quad \quad
a.a.s.
\end{eqnarray}
\end{corol}
\vspace{8pt}
\begin{proof}
see the Appendix.
\end{proof}
The throughput scaling law of $\log n$ is, by a factor of 4, larger
than the value obtained in \cite{Gowaikar05} in a centralized and
multihop scenario.

\section{Optimality Result}\label{UTMoptimal}

In this section, we provide an upper bound on the maximum throughput
of the wireless network in a Rayleigh fading environment. First, we
need the following lemma that provides a lower bound on the number
of active links.

\begin{lemma}\label{lowerboundnoactive}
In the optimum LAS, the number of active links, $k^*$, a.a.s.
satisfies
\begin{equation}
k^* \geq \dfrac{\log n}{\log \log n} \left( 1 + O\left(
\dfrac{1}{\log \log n} \right) \right).
\end{equation}
\end{lemma}

\begin{proof}
It can be shown that
\begin{equation}
g_{ii} \leq \log n + \varphi, \quad \forall i \in
\mathcal{N}_n,\quad a.a.s.,
\end{equation}
for any $\varphi=\omega(1)$. In the following, we assume
$\varphi=o(\log n)$. By ignoring the interference term and using the
above upper bound, the maximum throughput is upper bounded as
\begin{equation}
T^* \leq k^* \log \left( 1 + \dfrac{\log n + \varphi}{1/\rho}
\right), \quad a.a.s.
\end{equation}
Combining this upper bound with the lower bound in
(\ref{maxthroughputrayleigh1}), we obtain
\begin{eqnarray}
k^* & \geq & \dfrac{\log n + O(\log \log n)}{\log \left( 1 +
\dfrac{\log n + \varphi}{1/\rho} \right)}\\
& = & \dfrac{\log n}{\log \log n} \left( 1 + O\left( \dfrac{1}{\log
\log n} \right) \right),
\end{eqnarray}
where the equality is obtained by using $\varphi=o(\log n)$.
\end{proof}

\begin{theorem}\label{upperboundthroughput}
Consider a wireless network with $n$ links and i.i.d. random channel
coefficients drawn from an exponential distribution with mean
$\mu=1$. The maximum throughput over all sets of active links is
a.a.s. upper bounded as
\begin{equation}\label{throughputupperbound}
T^* \leq \log n +  \log \log n (1+  o (1)).
\end{equation}
\end{theorem}

\begin{proof}
For a randomly selected set of active links $\mathcal{A}$ with
$|\mathcal{A}|=k^*$, the interference term $I_i= \sum_{\substack{j \in \mathcal{A}\\
j \neq i}} g_{ji}$ in the denominator of (\ref{SINRdef}) has
$\chi^2(2k^*-2)$ distribution. Hence, we have
\begin{eqnarray}\label{ccdfSINR}
\nonumber \textrm{P}(\gamma_i > x) & = & \int_0^\infty \textrm{P}\left(\gamma_i > x | I_i=z \right) f_{I_i}(z) dz \\
\nonumber & = & \int_0^\infty e^{-x (1/\rho+z)} \dfrac{z^{k^*-2}e^{-z}}{(k^*-2)!}dz \\
& = & \dfrac{e^{-x/\rho}}{(1+x)^{k-1}}.\label{ccdfSINR}
\end{eqnarray}
Consequently, by using (\ref{ratedef}), we obtain
\begin{eqnarray}\label{rateccdf}
\nonumber \textrm{P}(r_i > x) & = & \textrm{P}(\gamma_i >
e^x-1)\\
& = & \dfrac{e^{-(e^x-1)/\rho}}{e^{(k-1)x}}.\label{rateccdf}
\end{eqnarray}
By defining $X_i=r_i+\dfrac{e^{r_i}-1}{\rho(k^*-1)}$ and using
(\ref{rateccdf}), it can be shown that $X_i$ is exponentially
distributed with mean $\frac{1}{k^*-1}$. On the other hand, from the
definition of $X_i$ it is clear that $X_i \geq r_i$. Thus, the
throughput $T(\mathcal{A})= \sum_{i \in \mathcal{A}} r_i $ is upper
bounded as
\begin{equation}
T(\mathcal{A}) \leq \sum_{i \in \mathcal{A}} X_i.
\end{equation}
Consequently, we have
\begin{eqnarray}
\textrm{P} \left( T(\mathcal{A}) > x \right) & \leq & \textrm{P}
\left(  \sum_{i \in \mathcal{A}} X_i > x \right)\\
& \stackrel{(a)}{=} & e^{-(k^*-1)x} \sum_{m=0}^{k^*-1} \frac{((k^*-1)x)^m}{m!}\\
& \stackrel{(b)}{<} & k^* e^{-(k^*-1)x} \frac{((k^*-1)x)^{k^*-1}}{(k^*-1)!}\\
& \stackrel{(c)}{\approx} & \sqrt{k^*} e^{-(k^*-1)(x-1)} x^{k^*-1},
\label{throughputccdfupper}
\end{eqnarray}
where (a) is because $\sum_{i \in \mathcal{A}} X_i$ has
$\chi^2(2k^*)$ distribution, (b) is because the maximum of the
summand terms occurs at $m=k-1$ for large enough $x$ \footnote{Since
we are seeking an upper bound on the throughput, $x$ is at least of
order $\log n$. This value is large enough to satisfy the mentioned
condition.}, and (c) is obtained by applying the Stirling's
approximation for the factorial, i.e., $m! \approx \sqrt{2 \pi m}m^m
e^{-m}$.

Assume $\mathcal{L}$ is the event that there exists at least one set
$\mathcal{A} \subseteq \mathcal{N}_n$ with $|\mathcal{A}|=k^*$ such
that $ T(\mathcal{A}) > x$. We have
\begin{eqnarray}
\textrm{p}(\mathcal{L}) &  \leq & \binom{n}{k^*} \textrm{P} \left(
T(\mathcal{A}) > x \right)\\
& < & \left( \frac{ne}{k^*} \right)^{k^*} \sqrt{k^*}
e^{-(k^*-1)(x-1)}
x^{k^*-1}\\
& < & \exp(\mathcal{E}(x,\,k^*)),\label{unconstrainedexistanceupper}
\end{eqnarray}
where the first inequality is due to the union bound, the second
inequality is due to (\ref{throughputccdfupper}) and the Stirling's
approximation, and $\mathcal{E}(x,\,k^*)$ is defined as
\begin{equation}
\mathcal{E}(x,\,k^*)=k^*(\log n - x - \log k^* + \log x + 2) +
\frac{1}{2} \log k^* + x.
\end{equation}
For $x=\log n +  \log \log n + 2 \log \log \log n$, we have
\begin{eqnarray}
\nonumber \mathcal{E}(x,\,k^*) & \approx & -k^* (2 \log \log \log n
+ \log k^*
- 2) + \dfrac{1}{2} \log k^* \\
& & + \log n + \log \log n (1+o(1)). \label{exponent}
\end{eqnarray}
Noting that the RHS of (\ref{exponent}) is a decreasing function in
$k^*$, we can replace $k^*$ by its lower bound from
Lemma~\ref{lowerboundnoactive} to obtain the upper bound
\begin{equation}\label{exponentupper}
\mathcal{E}(x,\,k^*) \leq - \dfrac{\log \log \log n}{\log \log n}
\log n (1+o(1)).
\end{equation}
Since the RHS of (\ref{exponentupper}) goes to $- \infty$ when $n
\to \infty$, from (\ref{unconstrainedexistanceupper}) we conclude
that $\textrm{p}(\mathcal{L}) \to 0$. This means, with probability
approaching 1, there does not exist any set $\mathcal{A}$ that
achieves a throughput larger than $x=\log n +  \log \log n (1+  o
(1))$. This completes the proof.
\end{proof}

It should be noted that an upper bound of $2 \log n$ has been
derived in \cite{Gowaikar05}. However, this larger upper bound is
obtained in a different scenario than ours; they consider a rate
constraint for the active links as well as the possibility of
transmitter-receiver assignment.

Comparison between the achievability result in
Corollary~\ref{optimumthresholdrayleigh} and the upper bound in
Theorem~\ref{upperboundthroughput} reveals the following result.

\begin{theorem}
Consider a wireless network with $n$ links and i.i.d. random channel
coefficients drawn from an exponential distribution with mean
$\mu=1$. Then, the maximum throughput a.a.s. scales like $\log n$.
Moreover, this maximum throughput scaling law is a.a.s. achieved by
the distributed TBLAS presented in Section~\ref{achievable}.
\end{theorem}

\section{Conclusion}\label{conclusion}
In this paper, the throughput of single-hop wireless networks with
on-off strategy is investigated in a fading environment. To obtain a
lower bound on the throughput, a decentralized link activation
strategy is proposed and analyzed for a general fading model. It is
shown that in the popular model of Rayleigh fading a throughput of
order $\log n$ is achievable, which is by a factor of four larger
than what was obtained in previous works with centralized methods
\cite{Gowaikar05}. Moreover, for the Rayleigh fading model, an upper
bound of order $\log n$ is obtained that shows the optimality of the
proposed link activation strategy.

\appendix[Proof of Corollary~\ref{optimumthresholdrayleigh}]

The optimum value of the threshold, $\Delta^*$, is the value that
maximizes the achievable throughput in (\ref{originalobjective}). As
it is seen, $T_a (\Delta)$ is a complicated function of $\Delta$.
However, since $\xi$ can grow as slow as desired, we can set $\xi=0$
to obtain a more tractable form for $T_a (\Delta)$ from which a
\emph{zero order approximation} of the solution is obtained. In the
next stage, we will improve the solution using this zero order
approximation.

\paragraph{Zero order approximation} By setting $\xi=0$, the objective
function in (\ref{originalobjective}) is transformed to
\begin{equation}
\hat{T}_a (\Delta)= n e^{-\Delta} \log \left ( 1+ \dfrac{\Delta}{n
e^{-\Delta}} \right ).
\end{equation}
Using the approximation $\log (1+x) \approx x -\frac{x^2}{2}$, the
above function can be approximated as
\begin{equation}\label{zeroobjectiveapprox}
\hat{T}_a (\Delta) \approx \Delta-   \dfrac{\Delta^2}{2  n
e^{-\Delta} }.
\end{equation}
By taking the derivative of this function, it can be shown that the
value of $\Delta$ that maximizes $\hat{T}_a (\Delta)$, satisfies the
equation
\begin{equation}\label{zeroequation1}
2ne^{-\Delta}=2\Delta+\Delta^2.
\end{equation}
Denoting the solution of this equation by $\Delta^*_{(0)}$, it can
be verified that
\begin{equation}\label{zeroorderapprox}
\Delta^*_{(0)}= \log n - 2 \log \log n + \log 2 + O \left(
\dfrac{\log \log n}{\log n} \right).
\end{equation}

\paragraph{First order approximation}
Using $\Delta^*_{(0)}$ in (\ref{zeroorderapprox}), the term
containing $\xi$ in (\ref{originalobjective}) is approximated
as\footnote{With a little abuse of notation, we have replaced
$\dfrac{\xi}{\sqrt{2}}$ by $\xi$. This is acceptable, because we are
only interested in the order of the term that $\xi$ introduces to
the solution.}
\begin{equation}
\xi \sqrt{ne^{-t}}= \xi \log n.
\end{equation}
Since $\psi$ can be chosen of order $o(\log n)$, it is negligible in
comparison with $\xi \log n$. Thus, the function to be maximized
takes the form
\begin{equation}
T_a (\Delta)= \left ( n e^{-\Delta} -  \xi \log n \right ) \log
\left ( 1+ \dfrac{\Delta}{n e^{-\Delta} -  \xi \log n} \right ).
\end{equation}
Assuming $\xi=o(\log \log n)$, and taking the same approach as for
obtaining $\Delta^*_{(0)}$, we obtain
\begin{equation}
\Delta^*_{(1)}= \log n - 2 \log \log n + \log 2 + \dfrac{4 \log \log
n}{\log n}+ O \left ( \dfrac{\xi}{\log n}   \right).
\end{equation}
This confirms the value of $\Delta^*$ stated in
Corollary~\ref{optimumthresholdrayleigh}. By substituting the value
of $\Delta^*$ in (\ref{originalobjective}), the achievable
throughput is obtained as mentioned in the lemma. The number of
active links and the rate-per-link are obtained by using the value
of $\Delta^*$ in (\ref{optimumdelta}) and (\ref{optimumsumrate}),
respectively.

\bibliographystyle{IEEEtran} 
\bibliography{references} 

\begin{thebibliography}{10}
\providecommand{\url}[1]{#1}
\csname url@rmstyle\endcsname
\providecommand{\newblock}{\relax}
\providecommand{\bibinfo}[2]{#2}
\providecommand\BIBentrySTDinterwordspacing{\spaceskip=0pt\relax}
\providecommand\BIBentryALTinterwordstretchfactor{4}
\providecommand\BIBentryALTinterwordspacing{\spaceskip=\fontdimen2\font plus
\BIBentryALTinterwordstretchfactor\fontdimen3\font minus
  \fontdimen4\font\relax}
\providecommand\BIBforeignlanguage[2]{{%
\expandafter\ifx\csname l@#1\endcsname\relax
\typeout{** WARNING: IEEEtran.bst: No hyphenation pattern has been}%
\typeout{** loaded for the language `#1'. Using the pattern for}%
\typeout{** the default language instead.}%
\else
\language=\csname l@#1\endcsname
\fi
#2}}

\bibitem{Gupta00}
P.~Gupta and P.~R. Kumar, ``The capacity of wireless networks,'' \emph{IEEE
  Trans. Information Theory}, vol.~46, no.~2, pp. 388--404, March 2000.

\bibitem{Gastpar05}
M.~Gastpar and M.~Vetterli, ``On the capacity of large gaussian relay
  networks,'' \emph{IEEE Trans. Information Theory}, vol.~51, no.~3, pp.
  765--779, March 2005.

\bibitem{Xie04}
L.-L. Xie and P.~R. Kumar, ``A network information theory for wireless
  communication: scaling laws and optimal operation,'' \emph{IEEE Trans.
  Information Theory}, vol.~50, no.~5, pp. 748--767, May 2004.

\bibitem{Leveque05}
O.~L\'ev\^eque and E.~Telatar, ``Information theoretic upper bounds on the
  capacity of large extended ad hoc wireless networks,'' \emph{IEEE Trans.
  Information Theory}, vol.~51, no.~3, pp. 858--865, March 2005.

\bibitem{Franceschetti07j}
M.~Franceschetti, O.~Dousse, D.~Tse, and P.~Thiran, ``Closing the gap in the
  capacity of wireless networks via percolation theory,'' \emph{IEEE Trans.
  Information Theory}, vol.~53, no.~3, pp. 1009--1018, March 2007.

\bibitem{Grossglauser02}
M.~Grossglauser and D.~N.~C. Tse, ``Mobility increases the capacity of ad hoc
  wireless networks,'' \emph{IEEE/ACM Trans. Networking}, vol.~10, no.~4, pp.
  477--486, August 2002.

\bibitem{Kulkarni04}
S.~R. Kulkarni and P.~Viswanath, ``A deterministic approach to throughput
  scaling in wireless networks,'' \emph{IEEE Trans. Information Theory},
  vol.~50, no.~6, pp. 1041--1049, June 2004.

\bibitem{Toumpis04}
S.~Toumpis and A.~J. Goldsmith, ``Large wireless networks under fading,
  mobility, and delay constraints,'' in \emph{Proc. IEEE Infocom}, vol.~1, Hong
  Kong, 2004, pp. 609--619.

\bibitem{Xue05}
F.~Xue, L.-L. Xie, and P.~R. Kumar, ``The transport capacity of wireless
  networks over fading channels,'' \emph{IEEE Trans. Information Theory},
  vol.~51, no.~3, pp. 834--847, March 2005.

\bibitem{Gowaikar05}
R.~Gowaikar, B.~Hochwald, and B.~Hassibi, ``Communication over a wireless
  network with random connections,'' \emph{IEEE Trans. Information Theory},
  vol.~52, no.~7, pp. 2857--2871, July 2006.

\bibitem{Etkinthesis}
R.~Etkin, ``Spectrum sharing: Fundamental limits, scaling laws, and
  self-enforcing protocols,'' Ph.D. dissertation, EECS Department, University
  of California, Berkeley, 2006.

\bibitem{Weber06}
S.~Webere, J.~G. Andrews, and N.~Jindal, ``Ad hoc networks: the effect of
  fading, power control, and fully-distributed scheduling,'' \emph{Submitted to
  IEEE Trans. Information Theory}, 2006.

\bibitem{Gesbert07rawnet}
D.~Gesbert and M.~Kountouris, ``Resource allocation in multicell wireless
  networks: Some capacity scaling laws,'' in \emph{Proc. Workshop on Resource
  Allocation in Wireless NETworks (RAWNET '07)}, 2007.

\bibitem{MyTechnical2}
M.~Ebrahimi, M.~A. Maddah-Ali, and A.~K. Khandani, ``Throughput scaling in
  decentralized single-hop wireless networks with fading channels,'' University
  of Waterloo, Tech. Rep. UW-ECE \#2006-13, 2006, available at
  http://www.cst.uwaterloo.ca/{pub-tech-rep.html}.

\end{thebibliography}

\end{document}